\documentclass[twocolumn,aps,superscriptaddress,showpacs]{revtex4}

\usepackage{amssymb}
\usepackage{amsmath}
\usepackage{graphicx}
\usepackage[normalem]{ulem}
\usepackage[dvips]{color}

\usepackage[normalem]{ulem}  % \sout{old text} for strikeout
\usepackage[dvips]{color} % For blue in-text comments and additions

\renewcommand\sout{\bgroup \color{red} \ULdepth=-.5ex \ULset}
\begin{document}
\title{Nuclear constraints on non-Newtonian gravity at femtometer scale}
\author{Jun Xu}\email{Jun.Xu@tamuc.edu}
\affiliation{Department of Physics and Astronomy, Texas A$\&$M
University-Commerce, Commerce, Texas 75429-3011, USA}
\author{Bao-An Li}\email{Bao-An.Li@tamuc.edu}
\affiliation{Department of Physics and Astronomy, Texas A$\&$M
University-Commerce, Commerce, Texas 75429-3011, USA}
\affiliation{School of Science, Xi'an Jiao Tong
University, Xi'an 710049, P.R. China}
\author{Lie-Wen Chen}\email{lwchen@sjtu.edu.cn}
\affiliation{Department of Physics, Shanghai Jiao Tong University,
Shanghai 200240, China}
\affiliation{Center of Theoretical Nuclear
Physics, National Laboratory of Heavy Ion Accelerator, Lanzhou
730000, China}
\author{Hao Zheng}\email{zh-i@sjtu.edu.cn}
\affiliation{Department of Physics, Shanghai Jiao Tong University,
Shanghai 200240, China}

\begin{abstract}
Effects of the non-Newtonian gravity on properties of finite nuclei
are studied by consistently incorporating both the direct and
exchange contribution of the Yukawa potential in the Hartree-Fock
approach using a well-tested Skyrme force for the strong
interaction. It is shown for the first time that the strength of the
Yukawa term in the non-Newtonian gravity is limited to
$\log(|\alpha|)<1.75/[\lambda(\rm fm)]^{0.54} + 33.6$ within the
length scale of $\lambda=1-10$ fm in order for the calculated properties of finite
nuclei not to be in conflict with accurate experimental data available.
\end{abstract}

\pacs{%21.65.-f, %Nuclear matter
      21.30.Fe, %Forces in hadronic systems and effective interactions
      04.50.-h, %Higher-dimensional gravity and other theories of gravity
      21.10.Gv  %Nucleon distributions and halo features
      14.70.Pw, %Other gauge bosons
      }

\maketitle

The quest to unify gravity with other fundamental forces is among
the most challenging scientific questions for the new century
\cite{11questions}. While it is generally assumed that non-relativistic gravity obeys Newton's Inverse-Square-Law (ISL)
for all distances greater than the Plank length of about $1.6\times
10^{-33}$ cm \cite{Hoyle01,Adel03,New09}, the Newtonian gravity has been
tested and various upper limits \cite{Fis99,Rey05,Uzan03} on the ISL
violation have been set down to only about 10 fm so far \cite{Kam08}.
In fact, motivated by the possible existence of both extra dimensions within
string/M theories and new particles in the supersymmetric extension
of the Standard Model, it has long been proposed that the
Newtonian gravitational potential between two objects of masses
$m_1$ and $m_2$ at positions $\vec{r}_1$ and $\vec{r}_2$ may be
modified to~\cite{Fuj71}
\begin{equation}\label{Fu}
V_{\rm grav}(\vec{r}_1,\vec{r}_2) =
-\frac{Gm_1m_2}{|\vec{r}_1-\vec{r}_2|}(1+\alpha
e^{-|\vec{r}_1-\vec{r}_2|/\lambda}),
\end{equation}
where $G$ is the gravitational constant, $\alpha$ and $\lambda$ are
the strength and length scale of the non-Newtonian Yukawa potential,
respectively. Moreover, several modified gravity theories including
the scalar-tensor-vector gravity~\cite{Wof06} and $f(R)$ gravity~\cite{Cap09}
lead to such a non-Newtonian potential at the weak-field limit.
While the main purpose of this work is to constrain the parameters of the Yukawa term
at femtometer scales, it is important to emphasize that this kind of non-Newtonian
potentials have been tested extensively against well-known observations at galaxy scales,
see, e.g., Ref. \cite{Nap12} for a recent review. In particular, we notice here that
the non-Newtonian potential of Eq.~(\ref{Fu}) has been used to successfully explain
both the flattening of the galaxy rotation curves away from the Kepler limit, see, e.g. Refs.~\cite{San84,Mof96}
and the Bullet Cluster 1E0657-558 observations in the absence of dark matter~\cite{Bro07}.
Moreover, it was shown that, unlike for massive particles, the motion of massless particles is not affected by the
Yukawa term. Consequently, all the lensing observables obtained with the non-Newtonian potential of Eq.~(\ref{Fu})
are equal to the ones known from General Relativity thanks to suitable cancelations in the post-Newtonian limit~\cite{Lub11,Sta12}.
Thus, to explain the gravitational lensing phenomenon, dark matter seems still needed with the non-Newtonian potential considered here.

In the one-boson-exchange picture, the Yukawa potential may come from
exchanging a light and weakly coupled spin-0 axion~\cite{Wil78} or
spin-1 $U$-boson~\cite{Fay80} corresponding respectively to an
attractive or repulsive potential, and it can be further written as
\begin{equation}
V_{\rm Y}(\vec{r}_1,\vec{r}_2) = \pm \frac{g^2}{4\pi} \frac{e^{-\mu
|\vec{r}_1-\vec{r}_2|}}{|\vec{r}_1-\vec{r}_2|},
\end{equation}
where $\mu=1/\lambda$ is the boson mass, and $g=\sqrt{4\pi|\alpha| G
m^2}$ is the boson-nucleon coupling constant with $m$ being the
nucleon mass. The exchanged boson may mediate the annihilation of
dark matter particles~\cite{Boe04a,Boe04b} and it is related to
several interesting new phenomena in particle physics and
cosmology~\cite{Fay07,Fay09}. It was also found recently that the
Yukawa term affects significantly the Equation of State (EOS) of
neutron-rich nuclear matter, thus properties of neutron stars (NSs),
e.g., their mass-radius relation, moment of inertia and the
core-crust transition density/pressue
\cite{Kri09,Wen09,Zhang11,Zhe11}. While observed properties of NSs
can be very well reproduced by combining the Yukawa term using
appropriate parameters with nuclear EOSs that otherwise failed to do
so by themselves \cite{Kri09,Wen09}, the Yukawa term is not
absolutely necessary because of the poorly known nature of the EOS
of dense neutron-rich nuclear matter. On the other hand, it is not
ruled out either. To our best knowledge, no quantitative constraint
exists on the strength of the Yukawa term at femtometer or shorter
length scale.

Nuclei are testing grounds of fundamental interactions among
nucleons in the femtometer range. Generally speaking, experimental measurements of
global properties of stable nuclei, such as their binding energies
and charge radii, are very well reproduced consistently within
about $2\%$ uncertainty using established nuclear many-body theories and our current knowledge about
the strong, weak and Coulomb forces without considering gravity.
Moreover, these properties are most strongly influenced
by the well known isoscalar part of the nuclear strong interaction
around the saturation density of nuclear matter. Thus, in order
to be consistent with current theoretical understanding and experimental observations in nuclear physics, effects of
gravity on properties of stable nuclei have to be less than about $2\%$.
Using this requirement and by consistently incorporating both
the direct and exchange contributions of gravity in the
Hartree-Fock (HF) approach using a well established Skyrme force for the
strong interaction, we show for the first time that the strength of
the Yukawa term is limited to $\log(|\alpha|)<1.75/[\lambda(\rm
fm)]^{0.54} + 33.6$ in the range of $\lambda=1-10$ fm. This provides
a reliable reference to test stringently non-Newtonian gravitational theories in
this previously unexplored region.

The contribution from the non-Newtonian Yukawa potential to the
potential energy of the nuclear system is
\begin{equation}\label{eub}
E_{\rm Y} = \frac{1}{2} \sum_{i,j} \langle ij|(1-P_r P_\sigma
P_\tau) V_{\rm Y} |ij\rangle,
\end{equation}
where $P_r$, $P_\sigma$, and $P_\tau$ are the space, spin, and
isospin exchange operator, respectively, and $|i\rangle$ is the
quantum state of the $i$th particle containing spatial, spin, and
isospin parts. The first term in Eq.~(\ref{eub}) is the direct
contribution and can be calculated from
\begin{equation}\label{EUBD}
E_{\rm Y}^D = \frac{1}{2} \int \rho(\vec{r}_1) \rho(\vec{r}_2)
\frac{g^2}{4\pi} \frac{e^{-\mu
|\vec{r}_1-\vec{r}_2|}}{|\vec{r}_1-\vec{r}_2|} d^3r_1 d^3r_2,
\end{equation}
where $\rho(\vec{r})=\sum_{i,\sigma,\tau} \phi_{\tau
i}^\star(\vec{r},\sigma)\phi_{\tau i}(\vec{r},\sigma)$ is the
nucleon density with $\phi_{\tau i}(\vec{r},\sigma)$ being the
spatial wave function of the $i$th particle with spin $\sigma$ and
isospin $\tau$. Using $P_\sigma =
(1+\vec{\sigma}_1\cdot\vec{\sigma}_2)/2$, with $\vec{\sigma}_{1(2)}$
being the Pauli operator acting on the first (second) term, the
second term in Eq.~(\ref{eub}) representing the exchange
contribution can be expressed as
\begin{eqnarray}
E_{\rm Y}^E &=& -\frac{1}{4} \sum_{\tau=n,p} \int
[\rho_\tau(\vec{r}_1,\vec{r}_2)\rho_\tau(\vec{r}_2,\vec{r}_1)
\notag\\
&+& \vec{\rho_\tau}(\vec{r}_1,\vec{r}_2) \cdot
\vec{\rho_\tau}(\vec{r}_2,\vec{r}_1)] \frac{g^2}{4\pi} \frac{e^{-\mu
|\vec{r}_1-\vec{r}_2|}}{|\vec{r}_1-\vec{r}_2|} d^3r_1 d^3r_2, \notag\\
\end{eqnarray}
where $\rho_\tau(\vec{r}_1,\vec{r}_2) = \sum_{i,\sigma}\phi_{\tau
i}^\star(\vec{r}_1,\sigma) \phi_{\tau i}(\vec{r}_2,\sigma)$ and
$\vec{\rho}_\tau(\vec{r}_1,\vec{r}_2) = \sum_i
\sum_{\sigma,\sigma^\prime}\phi_{\tau
i}^\star(\vec{r}_1,\sigma^\prime) \phi_{\tau i}(\vec{r}_2,\sigma)
\langle \sigma^\prime| \vec{\sigma}|\sigma \rangle$ are the
off-diagonal scalar and vector part of the density-matrix,
respectively. Introducing the coordinate transformation $\vec{r} =
(\vec{r}_1+\vec{r}_2)/2$ and $\vec{s} = \vec{r}_1-\vec{r}_2$, they
can be calculated from the density-matrix expansion
method~\cite{Neg72,Xu10b}
\begin{eqnarray}
&&\rho_\tau(\vec{r}+\frac{\vec{s}}{2},\vec{r}-\frac{\vec{s}}{2})
\rho_\tau(\vec{r}-\frac{\vec{s}}{2},\vec{r}+\frac{\vec{s}}{2})  \notag\\
&\approx& \rho_\tau^2(\vec{r}) \rho_{SL}^2(k_\tau s) + 2
\rho_\tau(\vec{r}) \rho_{SL}(k_\tau s) g(k_\tau s) s^2
\notag\\
&\times&\left[ \frac{1}{4} \nabla^2 \rho_\tau(\vec{r}) -
\tau_\tau(\vec{r}) + \frac{3}{5}k_\tau^2 \rho_\tau (\vec{r})
\right],\\
&&\vec{\rho}_\tau(\vec{r}+\frac{\vec{s}}{2},\vec{r}-\frac{\vec{s}}{2})
\approx \frac{i}{2}j_0(k_\tau s) \vec{s} \times
\vec{J}_\tau(\vec{r}).
\end{eqnarray}
In the above, $k_\tau = (3\pi^2 \rho_\tau)^{1/3}$ is the Fermi
momentum, $\tau_\tau(\vec{r})=\sum_{i,\sigma}|\nabla \phi_{\tau
i}(\vec{r},\sigma)|^2$ is the kinetic energy density, and
$\vec{J}_\tau(\vec{r})=-i\sum_i
\sum_{\sigma,\sigma^\prime}\phi_{\tau
i}^\star(\vec{r},\sigma^\prime) \nabla\phi_{\tau i}(\vec{r},\sigma)
\times \langle \sigma^\prime| \vec{\sigma}|\sigma \rangle$ is the
spin density. The $\rho_{SL}(k_\tau s)$ and $g(k_\tau s)$ can be
expressed in terms of the first- and the third-order spherical
Bessel function as $\rho_{SL}(k_\tau s) = 3j_1(k_\tau s)/(k_\tau s)
$ and $g(k_\tau s) = 35 j_3 (k_\tau s)/[2(k_\tau s)^3] $,
respectively.

Using the expressions from the density-matrix expansion, the potential energy density
from the exchange contribution of the Yukawa potential can thus be
written as
\begin{eqnarray}
H_{\rm Y}^E(\vec{r}) &=& \sum_{\tau=n,p} \{ A[\rho_\tau(\vec{r})] +
B[\rho_\tau(\vec{r})]\tau_\tau(\vec{r}) \notag\\
&+& C[\rho_\tau(\vec{r})] [\nabla \rho_\tau(\vec{r})]^2 +
\varphi[\rho_\tau(\vec{r})]J_\tau^2(\vec{r}) \},
\end{eqnarray}
where
\begin{eqnarray}
A(\rho_\tau) &=& - \frac{1}{4} \int \rho_\tau^2
\rho_{SL}^2(k_\tau s) V_{\rm Y}(s) d^3 s \notag\\
&-& \frac{3}{5} (3\pi^2)^{2/3} \rho_\tau^{5/3} B(\rho_\tau),\label{a}\\
B(\rho_\tau) &=& \frac{1}{2} \int \rho_\tau \rho_{SL}(k_\tau s)
g(k_\tau
s) s^2 V_{\rm Y}(s) d^3 s,\label{b}\\
C(\rho_\tau) &=& \frac{1}{4}\frac{dB(\rho_\tau)}{d\rho_\tau},\label{c}\\
\varphi(\rho_\tau) &=& -\frac{\pi}{6} \int j_0^2(k_\tau s) s^4
V_{\rm Y}(s) ds,
\end{eqnarray}
with $V_{\rm Y}(s) =  g^2e^{-\mu s}/(4\pi s)$.

To calculate properties of finite nuclei within the HF approach, we solve
the Schr\"{o}dinger equation
\begin{eqnarray}
&&\left[-\nabla \cdot \frac{\hbar^2}{2m_\tau^\star(\vec{r})} \nabla
+ U_\tau(\vec{r}) + \vec{W}_\tau (\vec{r}) \cdot (-i) (\nabla \times
\vec{\sigma})\right] \phi_{\tau i} \notag \\ &=& e_{\tau i}
\phi_{\tau i}
\end{eqnarray}
where $m_\tau^\star$, $U_\tau$ and $\vec{W}_\tau$ are the nucleon
effective mass, the single-particle potential, and the form factor
of the one-body spin-orbit potential, respectively. From the
variational principle, the Yukawa contribution to the
single-particle potential has a direct term $U_\tau^{\rm YD}$ and an
exchange term $U_\tau^{\rm YE}$, namely,
\begin{eqnarray}
U_\tau^{\rm Y} &=& U_\tau^{\rm YD} + U_\tau^{\rm YE} \\
U_\tau^{\rm YD} &=& \int \rho(\vec{r}^\prime) \frac{g^2}{4\pi}
\frac{e^{-\mu
|\vec{r}-\vec{r}^\prime|}}{|\vec{r}-\vec{r}^\prime|} d^3 r^\prime, \label{d} \\
U_\tau^{\rm YE} &=& \frac{dA(\rho_\tau)}{d\rho_\tau} +
\frac{dB(\rho_\tau)}{d\rho_\tau}\tau_\tau -
\frac{dC(\rho_\tau)}{d\rho_\tau} (\nabla \rho_\tau)^2  \notag\\
&-& 2C(\rho_\tau) \nabla^2 \rho_\tau +
\frac{d\varphi(\rho_\tau)}{d\rho_\tau} J_\tau^2. \label{ex1}
\end{eqnarray}
The Yukawa contribution to the form factor $\vec{W}_\tau^{\rm Y}$ is
\begin{equation}
\vec{W}_\tau^{\rm Y}=2\varphi(\rho_\tau)\vec{J}_\tau.
\label{ex2}
\end{equation}
In addition, the effective mass is modified by the Yukawa term according to
\begin{equation}
\frac{\hbar^2}{2m_\tau^\star} \rightarrow
\frac{\hbar^2}{2m_\tau^\star} + B(\rho_\tau). \label{ex3}
\end{equation}

Most available nuclear effective interactions have been tuned to
reproduce not only empirical properties of symmetric nuclear
matter at saturation density but also global properties of finite nuclei along the $\beta$ stability line.
These interactions differ mostly in their isovector parts and/or the effective
three-body forces used. In this work, we use properties, such as the
charge radii and binding energies of stable medium to heavy nuclei
that are not much affected by the still uncertain parts of the
strong nuclear interaction. Specifically, we use here the MSL0 parameter
set of a Skyrme-like force which has been tested extensively and was shown to describe the
charge radii and binding energies of finite nuclei very
well~\cite{Che10}.

\begin{figure}[ht]
\includegraphics[scale=0.8]{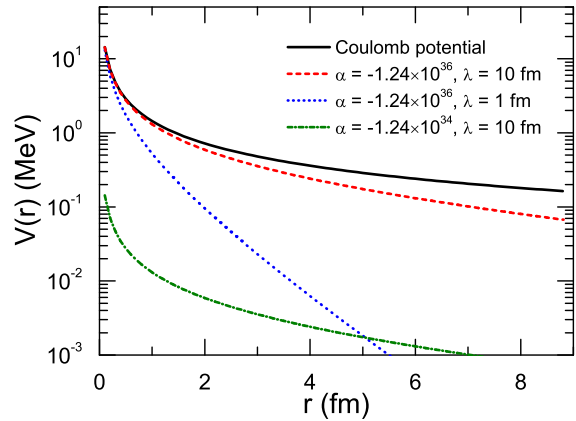}
\caption{(color online) A comparison of the Coulomb and Yukawa potentials
between two protons as a function of separation with different parameters.}\label{vr}
\end{figure}

Before exploring effects of gravity on properties of finite nuclei,
it is instructive to first compare the magnitudes of the Coulomb and
Yukawa potentials between two protons. Shown in Fig.~\ref{vr} are
the Yukawa potentials using $\alpha=-1.24\times10^{36}$ or
$-1.24\times10^{34}$, and $\lambda=1$ fm or 10 fm, respectively.
With $\alpha=-1.24\times10^{36}$, which leads to the coupling
constant $g^2/4 \pi = 1/137$, and a long interaction range of
$\lambda=10$ fm, the Yukawa potential approaches the Coulomb
potential at short distances, while it decreases faster towards long
distances for smaller values of $\lambda$.

\begin{figure}[ht]
\includegraphics[scale=0.8]{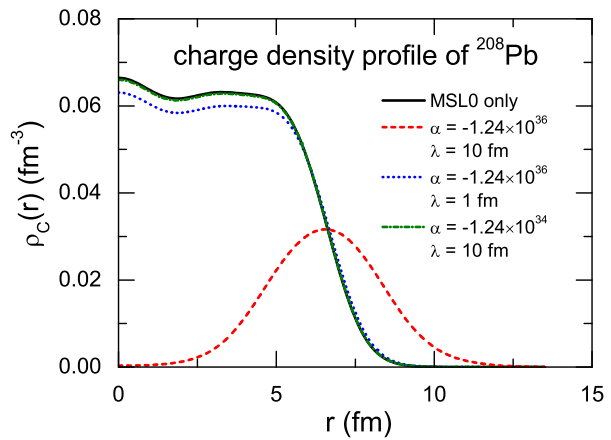}
\caption{(color online) Charge density profiles of $^{208}$Pb from
MSL0 with and without the Yukawa potential from different
parameters.}\label{den}
\end{figure}

As an example illustrating effects of the Yukawa potential on
properties of finite nuclei, shown in Fig.~\ref{den} are the charge
density profiles of $^{208}$Pb obtained within the HF approach using
the MSL0 interaction with different Yukawa parameters. With $\alpha$
on the order of $10^{34}$, the Yukawa potential has negligible
effects on the charge density profile. Increasing the magnitude of $\alpha$ to
$10^{36}$, effects of the Yukawa potential become significant especially for larger values of $\lambda$.

\begin{figure}[ht]
\hspace{-1cm}\includegraphics[scale=0.8]{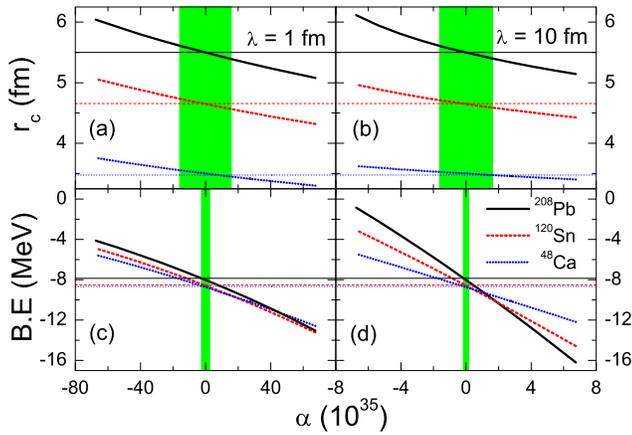}
\caption{(color online) Charge radius $r_c$ [(a), (b)] and binding
energy (B.E.) [(c), (d)] of $^{208}$Pb, $^{120}$Sn, and $^{48}$Ca as
functions of $\alpha$ at different length scales. The results at
$\alpha=0$ are those from the MSL0 interaction only. The green bands
are constrained $\alpha$ values from taking $2\%$ as the uncertainty of the
charge radii and binding energies. }\label{properties}
\end{figure}

Next we investigate more systematically how the Yukawa potential
affects the charge radii $r_c$ and binding energies (B.E.) of three
typical medium to heavy nuclei by varying continuously the strength
parameter $\alpha$. Shown in Fig.~\ref{properties} are the results
for $^{208}$Pb, $^{120}$Sn, and $^{48}$Ca with $\lambda=1$ fm and 10
fm, respectively. The horizontal lines are the mean values of the
experimental data~\cite{Aud03,Ang04}. As the Yukawa potential varies
from being repulsive (with negative values for $\alpha$) to attractive, the charge radius decreases and
the nuclei becomes more bound. The effect is roughly proportional to
the value of $\alpha$. Moreover, the effects are larger for heavier
nuclei due to the finite-range nature of the Yukawa potential.

\begin{figure}[t]
\vspace{0.5cm}
\includegraphics[scale=0.8]{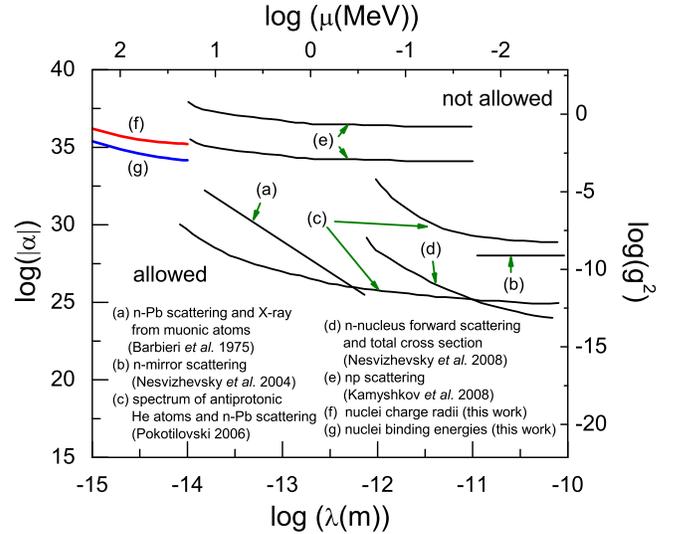}
\caption{(color online) The constraints of $\alpha$ for
$\lambda=1-10$ fm from charge radii and binding energies of heavy
nuclei and those at longer distances extracted from analyzing
neutron scattering experiments in Refs.
\protect{\cite{Bar75,Nes04,Pok06,Kam08,Nes08}.}}\label{constraint}
\end{figure}

As mentioned earlier, essentially all existing nuclear effective interactions
can describe the experimental charge radii and binding energies of
medium to heavy nuclei within about $2\%$. The latter is thus the
largest room available to accommodate effects of gravity. As shown
by the green bands in Fig.~\ref{properties}, this then allows us to
set a nuclear upper limit on the strength of the Yukawa potential in
the length range of 1 to 10 fm. Here we use the Yukawa potential effects on
$^{208}$Pb which is more strongly affected by gravity. More
specifically, the constraint from the binding energies can be
parameterized as $\log(|\alpha|)<1.75/[\lambda(\rm fm)]^{0.54} +
33.6$ while that from the charge radii can be written as
$\log(|\alpha|)<1.18/[\lambda(\rm fm)]^{0.79} + 35.0$ for
$\lambda=1-10$ fm. Since the binding energy is more sensitive to the
strength of Yukawa potential, the first constraint is more stringent
and can be used as the nuclear upper limit. This constraint thus
limits the coupling constant of the exchanged bosons with nucleons
to $\log(g^2)<0.10 [\mu (\rm MeV)]^{0.54}-3.53$ if the mass of the
bosons is between $20$ MeV and $200$ MeV. Previously, the shortest
range probed is above 10 fm in neutron scattering
experiments~\cite{Bar75,Nes04,Pok06,Kam08,Nes08}. For a comparison,
shown in Fig.~\ref{constraint} are the various upper limits on the
magnitude of $\alpha$ in the range of $10^{-15}$ to $10^{-10}$ m.
The constraint extracted from the present work extends to the
previously unexplored region of $1-10$ fm. Overall, there is a clear
trend that the allowed deviation from Newtonian gravity increases as
the interaction range decreases, reflecting the increasing
difficulties of the measurement.

In summary, effects of the non-Newtonian gravity on properties of
finite nuclei were studied within the Hartree-Fock approach
incorporating a well-tested Skyrme force for the strong interaction
and the non-Newtonian gravitational potential. For the first time, the
strength of the Yukawa term is limited to
$\log(|\alpha|)<1.75/[\lambda(\rm fm)]^{0.54} + 33.6$ within the
length range of $\lambda=1-10$ fm in order for the calculated properties of finite nuclei
not to be in conflict with the very accurate experimental data available.
This constraint serves as a useful reference in constraining properties of weakly-coupled gauge bosons
and further explorations of possible extra dimensions at femtometer scale.

\begin{acknowledgments}
This work was supported in part by the US National Science grants
PHY-0757839 and PHY-1068022, the National Aeronautics and Space
Administration under grant NNX11AC41G issued through the Science
Mission Directorate, the National Natural Science Foundation of
China under Grant Nos. 10975097 and 11135011, Shanghai Rising-Star
Program under grant No. 11QH1401100, ¡°Shu Guang¡± project supported
by Shanghai Municipal Education Commission and Shanghai Education
Development Foundation, the Program for Professor of Special
Appointment (Eastern Scholar) at Shanghai Institutions of Higher
Learning, and the National Basic Research Program of China (973
Program) under Contract No. 2007CB815004.
\end{acknowledgments}

\end{document}